\def\alf1{ {\alpha\over\pi} }

\documentclass{PoS}
\usepackage{amsmath}
\usepackage{amssymb}
\usepackage{epsfig}
\usepackage{euscript}
\usepackage{fancybox}
\usepackage{cite}
\usepackage{epic}

\def\mathswitchr#1{\relax\ifmmode{\mathrm{#1}}\else$\mathrm{#1}$\fi}

\newcommand {\pslash}{\hbox{$\not\hbox{\kern-2.3pt $p$}$}}

\title{Planck Scale Cosmology and Asymptotic Safety in Resummed Quantum Gravity}

\ShortTitle{Planck Scale Cosmology and Asymptotic Safety in Resummed Quantum Gravity}

\author{\speaker{B.F.L. Ward}%
         \thanks{Work supported in part by NATO grant PST.CLG.980342.}\\
        Baylor University\\
        E-mail: \email{BFL\_Ward@baylor.edu}}

\abstract{In Weinberg's asymptotic safety approach, 
a finite
dimensional critical surface for a UV stable fixed point 
generates a theory of
quantum gravity with a finite number of physical parameters. We argue 
that, in an extension of Feynman's original formulation of the theory,
we recover this fixed-point UV
behavior from an exact re-arrangement of the respective perturbative series.
Our results are consistent with the exact field space
Wilsonian renormalization group results of 
Reuter {\it et al.} and with recent Hopf-
algebraic Dyson-Schwinger renormalization theory results of Kreimer. 
We obtain the first "first principles" predictions of 
the dimensionless gravitational
and cosmological constants and our results support 
the Planck scale cosmology of Bonanno and Reuter. We conclude with an
estimate for the currently observed value of the cosmological 
constant.\\
\vskip0.5cm
\centerline{BU-HEPP-10-06, ~Nov., 2010}}

\FullConference{35th International Conference of High Energy Physics\\
                 July 22-28, 2010\\
                 Paris, France}

\begin{document}
\section{Introduction}\label{intro}
In Ref.~\cite{wein1}, Weinberg suggested that the general theory of relativity may have a non-trivial UV fixed point, with a finite dimensional critical surface
in the UV limit, so that it would be asymptotically safe with an S-matrix
that depends on only a finite number of observable parameters. 
In Refs.~\cite{reutera,litim,perc}, strong evidence has been calculated
using Wilsonian~\cite{kgw} field-space exact renormalization group methods to support asymptotic safety for the Einstein-Hilbert theory.
We have shown in Refs.~\cite{bw1,bw2i} that the extension of the amplitude-based, exact resummation theory of Ref.~\cite{yfs} to the Einstein-Hilbert theory
(we call the extension resummed quantum gravity) leads to UV fixed-point behavior for the dimensionless
gravitational and cosmological constants, but with the bonus that the resummed theory is actually UV finite.
More evidence for asymptotic safety for quantum gravity has been calculated using causal dynamical triangulated lattice methods in Ref.~\cite{ambj}\footnote{We also note that the model in Ref.~\cite{horva} realizes many aspects
of the effective field theory implied by the anomalous dimension of 2 at the
UV-fixed point but it does so at the expense of violating Lorentz invariance.}.
There is no known inconsistency between our analysis
and Refs.~\cite{reutera,litim,perc,ambj}.
Our results are also consistent with the results on leg renormalizability of quantum gravity in Refs.~\cite{kreimer}.
Contact with experiment is now in order.\par
Specifically, in Ref.~\cite{reuter1}, it has been argued that the approach in Refs.~\cite{reutera,litim,perc} 
to quantum gravity
may provide a realization\footnote{The attendant scale choice $k\sim 1/t$ used in Refs.~\cite{reuter1} was also proposed in Ref.~\cite{sola1}.} of the successful
inflationary model~\cite{guth,linde} of cosmology
without the need of the inflaton scalar field: the attendant UV fixed point solution
allows one to develop Planck scale cosmology that joins smoothly onto
the standard Friedmann-Walker-Robertson classical descriptions so
that one arrives at a quantum mechanical 
solution to the horizon, flatness, entropy
and scale free spectrum problems. In Ref.~\cite{bw2i}, using
the 
resummed quantum gravity theory~\cite{bw1}, 
we recover the properties as used in Refs.~\cite{reuter1} 
for the UV fixed point with ``first principles''
predictions for the fixed point values of
the respective dimensionless gravitational and cosmological constants. 
Here, we carry the analysis one step further and arrive at a prediction for the observed cosmological constant $\Lambda$ in the
context of the Planck scale cosmology of Refs.~\cite{reuter1}.
We comment on the reliability of the result as well, as it will be seen
already to be relatively close to the observed value~\cite{cosm1}.
More such
reflections, as they relate to an experimentally testable union of the original ideas of Bohr and Einstein, will be taken up
elsewhere~\cite{elswh}.\par
The discussion is organized as follows. In the next section we review 
the Planck scale cosmology presented 
in Refs.~\cite{reuter1}. In Section 3 we review our results~\cite{bw2i} 
for the dimensionless gravitational and cosmological constants
at the UV fixed point. In Section 4, we combine the Planck scale cosmology 
scenario~\cite{reuter1} with our results to predict 
the observed value of 
the cosmological constant $\Lambda$.
\par
\section{Planck Scale Cosmology}
More precisely, we recall the Einstein-Hilbert 
theory
\begin{equation}
{\cal L}(x) = \frac{1}{2\kappa^2}\sqrt{-g}\left( R -2\Lambda\right)
\label{lgwrld1a}
\end{equation} 
where $R$ is the curvature scalar, $g$ is the determinant of the metric
of space-time $g_{\mu\nu}$, $\Lambda$ is the cosmological
constant and $\kappa=\sqrt{8\pi G_N}$ for Newton's constant
$G_N$. Using the phenomenological exact renormalization group
for the Wilsonian~\cite{kgw} coarse grained effective 
average action in field space, the authors in Ref.~\cite{reuter1} 
have argued that
the attendant running Newton constant $G_N(k)$ and running 
cosmological constant
$\Lambda(k)$ approach UV fixed points as $k$ goes to infinity
in the deep Euclidean regime: 
$k^2G_N(k)\rightarrow g_*,\; \Lambda(k)\rightarrow \lambda_*k^2$
for $k\rightarrow \infty$.\par
The contact with cosmology then proceeds as follows. Using a phenomenological
connection between the momentum scale $k$ characterizing the coarseness
of the Wilsonian graininess of the average effective action and the
cosmological time $t$, $k(t)=\frac{\xi}{t}$ for $\xi>0$, the authors
in Refs.~\cite{reuter1} show that the standard cosmological
equations admit of the following extension:
$(\frac{\dot{a}}{a})^2+\frac{K}{a^2}=\frac{1}{3}\Lambda+\frac{8\pi}{3}G_N\rho,\;\dot{\rho}+3(1+\omega)\frac{\dot{a}}{a}\rho=0,\;\dot{\Lambda}+8\pi\rho\dot{G_N}=0,\;G_N(t)=G_N(k(t)),\;\text{and}\;\Lambda(t)=\Lambda(k(t))$
for the density $\rho$ and scale factor $a(t)$
with the Robertson-Walker metric representation as
$ds^2=dt^2-a(t)^2\left(\frac{dr^2}{1-Kr^2}+r^2(d\theta^2+\sin^2\theta d\phi^2)\right)$
so that $K=0,1,-1$ correspond respectively to flat, spherical and
pseudo-spherical 3-spaces for constant time t.  The equation of state
is $p(t)=\omega \rho(t)$
where $p$ is the pressure. 
\par
Using the UV fixed points for $g_*$ and
$\lambda_*$, the authors in Refs.~\cite{reuter1}
show that the extended cosmological system given above admits, for $K=0$,
a solution in the Planck regime where $0\le t\le t_{\text{class}}$, with
$t_{\text{class}}$ a ``few'' times the Planck time $t_{Pl}$, which joins
smoothly onto a solution in the classical regime, $t>t_{\text{class}}$,
which coincides with standard Friedmann-Robertson-Walker phenomenology
but with the horizon, flatness, scale free Harrison-Zeldovich spectrum,
and entropy problems all solved purely by Planck scale quantum physics.
We now review the results in Refs.~\cite{bw2i} for these
UV limits
and show how to use them to predict the current value of $\Lambda$. 
\par
\section{$g_*$ and $\lambda_*$ in Resummed Quantum Gravity}
We start with the prediction for $g_*$, which we already presented in Refs.~\cite{bw1,bw2i}. 
We have shown in Refs.~\cite{bw1} that the large virtual IR effects
in the respective loop integrals for 
the scalar propagator in quantum general relativity 
can be resummed to the {\em exact} result
$i\Delta'_F(k)|_{\text{resummed}} =  \frac{ie^{B''_g(k)}}{(k^2-m^2-\Sigma'_s+i\epsilon)}$
for
$B''_g(k)=\frac{\kappa^2|k^2|}{8\pi^2}\ln\left(\frac{m^2}{m^2+|k^2|}\right)$,
where this form holds for the UV regime, so that 
the resummed scalar propagator
falls faster than any power of $|k^2|$. An analogous result~\cite{bw1} holds
for m=0. As $\Sigma'_s$, the residual self-energy function, starts in ${\cal O}(\kappa^2)$,
we may drop it in calculating one-loop effects. It follows that,
when the respective analogs of $i\Delta'_F(k)|_{\text{resummed}}$
are used for the
elementary particles, all quantum gravity loop 
corrections are UV finite~\cite{bw1}. 
%
\par
When we use our resummed propagator results, 
as extended to all the particles
in the SM Lagrangian and to the graviton itself,
%
the denominator 
of the graviton propagator becomes~\cite{bw1} ($M_{Pl}$ is the Planck mass)
$q^2+\Sigma^T(q^2)+i\epsilon\cong q^2-q^4\frac{c_{2,eff}}{360\pi M_{Pl}^2}$,
for
$c_{2,eff}=\sum_{\text{SM particles j}}n_jI_2(\lambda_c(j))
         \cong 2.56\times 10^4$
with $I_2$ given in Refs.~\cite{bw1}
and with $\lambda_c(j)=\frac{2m_j^2}{\pi M_{Pl}^2}$. 
$n_j$ is the number of effective degrees of 
freedom~\cite{bw1} of particle $j$ of mass $m_j$. 
We take the SM
masses as explained in Refs.~\cite{bw1,bw2i} following Refs.~\cite{neut,pdg2002,lewwg,cosm1}.
We also note that from Ref.\cite{tHvelt1} it also follows
that the value of $n_j$ for the graviton and its attendant ghost is $42$.
We thus identify (we use $G_N$ for $G_N(0)$)
$G_N(k)=G_N/(1+\frac{c_{2,eff}k^2}{360\pi M_{Pl}^2})$ so that
$g_*=\lim_{k^2\rightarrow \infty}k^2G_N(k^2)=\frac{360\pi}{c_{2,eff}}\cong 0.0442$, 
a pure property of the known world.\par
Turning now to $\lambda_*$, we use 
Einstein's equation 
$G_{\mu\nu}+\Lambda g_{\mu\nu}=-\kappa^2 T_{\mu\nu}$ 
in a standard notation where $G_{\mu\nu}=R_{\mu\nu}-\frac{1}{2}Rg_{\mu\nu}$,
$R_{\mu\nu}$ is the contracted Riemann tensor, and
$T_{\mu\nu}$ is the energy-momentum tensor. Working with
the representation $g_{\mu\nu}=\eta_{\mu\nu}+2\kappa h_{\mu\nu}$
for the flat Minkowski metric $\eta_{\mu\nu}=\text{diag}(1,-1,-1,-1)$
we may isolate $\Lambda$ in Einstein's 
equation by evaluating
its VEV (vacuum expectation value). 
For any bosonic quantum field $\varphi$ we use
the point-splitting definition (here, :~~: denotes normal ordering)
$\varphi(0)\varphi(0)=\lim_{\epsilon\rightarrow 0}\varphi(\epsilon)\varphi(0)
=\lim_{\epsilon\rightarrow 0} T(\varphi(\epsilon)\varphi(0))
=\lim_{\epsilon\rightarrow 0}\{ :\varphi(\epsilon)\varphi(0): + <0|T(\varphi(\epsilon)\varphi(0))|0>\}$
where the limit is taken with time-like $\epsilon\equiv(\epsilon,\vec{0})\rightarrow (0,0,0,0)\equiv 0$ respectively.  
A scalar then makes the contribution~\cite{bw1} to $\Lambda$ given by\footnote{We note the
use here in the integrand of $2k_0^2$ rather than the $2(\vec{k}^2+m^2)$ in Ref.~\cite{bw2i}, to be
consistent with $\omega=-1$~\cite{zeld} for the vacuum stress-energy tensor.}
$\Lambda_s=-8\pi G_N$\\
$\frac{\int d^4k}{2(2\pi)^4}\frac{(2k_0^2)e^{-\lambda_c(k^2/(2m^2))\ln(k^2/m^2+1)}}{k^2+m^2}
\cong -8\pi G_N[\frac{1}{G_N^{2}64\rho^2}]$,
where $\rho=\ln\frac{2}{\lambda_c}$, 
 and a Dirac fermion
contributes~\cite{bw1} $-4$ times  $\Lambda_s$ to
$\Lambda$. 
The deep UV limit of $\Lambda$ then becomes
$\Lambda(k) \operatornamewithlimits{\longrightarrow}_{k^2\rightarrow \infty} k^2\lambda_*$,\\
$\lambda_*=-\frac{c_{2,eff}}{2880}\sum_{j}(-1)^{F_j}n_j/\rho_j^2
\cong 0.0817$
where $F_j$ is the fermion number of $j$ and $\rho_j=\rho(\lambda_c(m_j))$.
We see again that 
$\lambda_*$ 
is
a pure prediction of our known world -- $\lambda_*$ would vanish
in an exactly supersymmetric theory.
Our results for $(g_*,\lambda_*)$ agree qualitatively with those in Refs.~\cite{reuter1}.
\par
\section{An Estimate of $\Lambda$}
To estimate the value of $\Lambda$ today, we take the normal-ordered form of Einstein's equation, $:G_{\mu\nu}:+\Lambda :g_{\mu\nu}:=-\kappa^2 :T_{\mu\nu}:$.
The coherent state representation of the thermal density matrix then gives
the Einstein equation in the form of thermally averaged quantities with
$\Lambda$ given by our result above in lowest order. 
Taking the transition time between the Planck regime and the classical Friedmann-Robertson-Walker regime at $t_{tr}\sim 25 t_{Pl}$ from Refs.~\cite{reuter1},
we introduce$\rho_\Lambda(t_{tr})\equiv\frac{\Lambda(t_{tr})}{8\pi G_N(t_{tr})}
         =\frac{-M_{Pl}^4(k_{tr})}{64}\sum_j\frac{(-1)^Fn_j}{\rho_j^2}$
and use the arguments in Refs.~\cite{branch-zap} ($t_{eq}$ is the time of matter-radiation equality) to get the 
first principles estimate, from the method of the operator field,
$\rho_\Lambda(t_0)\cong \frac{-M_{Pl}^4(1+c_{2,eff}k_{tr}^2/(360\pi M_{Pl}^2))^2}{64}\sum_j\frac{(-1)^Fn_j}{\rho_j^2}
         [ \frac{t_{tr}^2}{t_{eq}^2} \times (\frac{t_{eq}^{2/3}}{t_0^{2/3}})^3] \cong \frac{-M_{Pl}^2(1.0362)^2(-9.197\times 10^{-3})}{64}\frac{(25)^2}{t_0^2}\cong$\\ $ (2.400\times 10^{-3}eV)^4$
where we take the age of the universe to be $t_0\cong 13.7\times 10^9$ yrs. 
In the latter estimate, the first factor in the square bracket comes from the period from
$t_{tr}$ to $t_{eq}$ (radiation dominated) and the second factor
comes from the period from $t_{eq}$ to $t_0$ (matter dominated)
\footnote{The method of the operator field forces the vacuum energies to follow the same scaling as the non-vacuum excitations.}.
This estimate should be compared with the experimental result~\cite{cosm1}\footnote{See also Ref.~\cite{sola2} for an analysis that suggests a value for $\rho_\Lambda(t_0)$ that is qualitatively similar to this experimental result.}
$\rho_\Lambda(t_0)|_{\text{expt}}\cong (2.368\times 10^{-3}eV(1\pm 0.023))^4$.
\par
To sum up, our estimate, while it is definitely encouraging, 
is not a precision prediction,
as possible hitherto unseen degrees of freedom have not been included and $t_{tr}$ is not precise, yet. -- We thank Profs. L. Alvarez-Gaume and W. Hollik for the support and kind
hospitality of the CERN TH Division and the Werner-Heisenberg-Institut, MPI, Munich, respectively, where a part of this work was done.\par

\end{document}